\documentclass[11pt]{amsart}
\usepackage{amssymb}
\newcommand{\F}{\mathbb{F}}
\title[Defeating the Kalka--Teicher--Tsaban attack]{Defeating the Kalka--Teicher--Tsaban linear algebra attack on
the Algebraic Eraser}
\author{Dorian Goldfeld}
\address{Department of Mathematics\\
Columbia University\\
New York, NY 10027}

\email{goldfeld@columbia.edu}

\author{Paul E. Gunnells}

\address{Department of Mathematics and Statistics\\
University of Massachusetts\\
Amherst, MA 01003-9305}

\email{gunnells@math.umass.edu}
\theoremstyle{definition}
\newtheorem*{remark}{Remark}

\begin{document}
\begin{abstract}
The \emph{Algebraic Eraser} (AE) is a public key protocol for sharing
information over an insecure channel using commutative and
noncommutative groups; a concrete realization is given by
\emph{Colored Burau Key Agreement Protocol} (CBKAP).  In this paper,
we describe how to choose data in CBKAP to thwart an attack by
Kalka--Teicher--Tsaban.
\end{abstract}

\date{Nov. 9, 2011}

\maketitle

\section{Introduction} 
The \emph{Algebraic Eraser} (AE), due to
Anshel--Anshel--Goldfeld--Lemieux \cite{ae}, is a public key protocol
for sharing information over an insecure channel using commutative and
noncommutative groups.  The \emph{Colored Burau Key Agreement
Protocol} (CBKAP) is a concrete realization of the AE based on the
braid group and finite general linear groups.  The AE and CBKAP have
been proposed as a public key protocol suitable for use in
low-resource environments, such as passive RFID systems and
remote-sensing networks.

In \cite{ktt} Kalka--Teicher--Tsaban describe an attack on CBKAP based
on probabilistic group theory that tries to recover part of the
private data in CBKAP.  This data consists of two matrices $n_{a}$,
$n_{b}$ in a large finite general linear group.
Kalka--Teicher--Tsaban explain---under the assumption that $n_{a}$ and
$n_{b}$ are chosen according to a certain probability
distribution---how to detect nontrivial relations that $n_{a}$,
$n_{b}$ must satisfy.  This then limits the spaces in which $n_{a},
n_{b}$ live so that searching for them is feasible.

In this short note, we explain a simple technique for choosing
$n_{a}$, $n_{b}$ that defeats this attack.

\section{The Algebraic Eraser Key Agreement Protocol and CBKAP} 

Following \cite{ae}, we describe a protocol that allows two users
(Alice and Bob) to create a shared secret over a public channel. The
Algebraic Eraser protocol is built from the tuple
\[
(G,M,N,\Pi, E,A,B,N_{A},N_{B}),
\]
where the \emph{publicly known} elements are as follows:
\begin{itemize}
\item $G$ is a group, with identity element $e$.
\item $M, N$ are two monoids.  The monoid $M$ has a left $G$-action
denoted by $(g,m)\mapsto ^{g}m$.  We denote the operation in $M$ by a
dot: $\cdot$.  We denote the semidirect product of $M$ and $G$ by $M
\rtimes G$, and write the binary operation using $\circ$:
\[
(m_1,g_1) \circ (m_2,g_2) = \left(m_1\cdot\, ^gm_2, \; g_1g_2\right)
\]
for all $(m_1,g_1), (m_2, g_2) \in M\times G.$
\item $\Pi \colon M\rightarrow N$ is a monoid homomorphism.
\item $E$ is a function $E\colon (N\times G) \times (M\rtimes G) \to N
\times G$, called {\it $E$-multiplication}, defined as follows. For
all $(n,g) \in N\times G$ and all $(m, g') \in (M\rtimes G)$ we
put 
\[
E((n,g), (m, g')) := \left(n\cdot \Pi\left(^gm\right),
gg'\right) \in N\times G.
\]
We denote $E$-multiplication by a star: $E((n,g), (m, g')) = (n,g)*
(m,g')$.

\item $A,B \subset M\rtimes G$ are two $E$-commuting submonoids. Here by $E$-commuting we mean 
\[
\left(\Pi(a),
\; g_a\right) * (b, g_b) = \left(\Pi(b),\; g_b\right)* (a, g_a)
\]
holds for
all $(a, g_a) \in A, \; (b, g_b) \in B$.
\item Two commuting submonoids $N_A, N_B \subset N.$ 
\end{itemize}

Now we describe how this data is used to form the    \emph{AE Key Agreement Protocol}. 
The submonoids $A, N_A$ are assigned to Alice, while $B, N_B$ are
assigned to Bob.  Alice chooses \emph{private keys}
\[
n_a \in N_A, \quad (a_1,g_{a_1}), \ldots, (a_k, g_{a_k}) \in A
\]
and then builds the \emph{public key} 
\[
p_A = (n_a, e) * (a_1, g_{a_1}) * \cdots * (a_k, g_{a_k}) \in N\times G.
\]
Similarly, Bob chooses private keys 
\[
n_b \in N_B, \quad (b_1,g_{b_1}), \ldots, (b_\ell, g_{b_\ell}) \in B
\]
and the public key 
\[
p_B = (n_b, e) * (b_1, g_{b_1}) * \cdots * (b_\ell, g_{b_\ell}) \in N\times G.
\]
Given this data, Alice and Bob can then each compute one side of the
following equation, which constitutes the  shared secret of the protocol:
\[
{(n_b, e) \cdot p_A *(b_1, g_{b_1}) * \cdots *(b_\ell, g_{b_\ell})  = (n_a, e) \cdot p_B *(a_1, g_{a_1}) * \cdots *(a_\ell, g_{a_\ell}).}
\]   
We note that, in practice, all data in the protocol would be assigned
to Alice and Bob by a trusted third party (TTP).

We now describe the \emph{Colored Burau Key Agreement Protocol}, an
explicit instance of the AE.  Choose $n\geq 8$ even and let $t =
(t_{1},\dotsc ,t_{n})$ be a tuple of variables.  Define matrices
$x_{i} (t)$ by
\[
x_{1} (t) = \left(\begin{array}{cccc}
-t_{1}&1&&\\
&1&&\\
&&\ddots&\\
&&&1
\end{array} \right),
\]
and for $i=2,\dotsc ,n-1$ by
\[
x_{i} (t) = \left(\begin{array}{ccccc}
1&&&&\\
&\ddots&&&\\
&t_{i}&-t_{i}&1&\\
&&&\ddots&\\
&&&&1
\end{array} \right).
\]
Fix a finite field $\F$.  The matrices $x_{i} (t)$ generate a subgroup
\[
M\subset  GL (n, \F (t_{1},\dotsc ,t_{n-1})).
\]
Let $G=S_{n}$, the symmetric group on $n$ letters, act on the $t_{i}$
by permutations, and let $s_{i}\in S_{n}$ be the simple transposition
$(i,i+1)$.  Then the pairs $\{(x_{i} (t), s_{i}) \}$ then generate the
semidirect product $M\rtimes S_{n}$ inside $GL (n, \F (t_{1},\dotsc
,t_{n-1})) \rtimes S_{n}$. 
Let $N=GL (n,\F)$ and choose $n-1$ nonzero elements $\tau_{i}\in N$.
The assignment $t_{i}\mapsto \tau_{i}$ defines a map $\Pi \colon
M\rightarrow N$.

To complete the description of CBKAP, we only need to specify the
commuting monoids $A,B\subset M$ and the $E$-commuting monoids $N_{A},
N_{B}\subset N$.  For the former, we can take $A$ (respectively, $B$)
to be the subgroup generated by the first (resp., last) $(n-2)/2$
matrices $x_{i} (t)$.  For the latter, we can fix a matrix $m\in N$
and then define $N_{A} = N_{B} = \F [m]$, where the latter means all
polynomials in $m$ with coefficients in $\F$ that lie in $N$.  How one
chooses $m$ will be explained below in \S\ref{defeat}.

\section{The Kalka-Teicher-Tsaban Attack} 

In [KTT] a practical linear algebraic attack on the AE is
developed. The attacker (called Eve) attempts to find Bob's first
private key $n_b \in N_B$. The attack goes as follows. To
attack the AE key agreement protocol, Eve creates a spurious
element $$(\alpha, e) \in A \subset M\rtimes G.$$ Then $(\alpha, e)$
$E$-commutes with every element in $B$. In particular it $E$-commutes
with $$(\beta, g) := (b_1, g_{b_1})\circ \cdots \circ (b_\ell,
g_{b_\ell}),$$ given by taking the semidirect product of Bob's second
private keys. It follows that
\begin{align*}
\left(\Pi(\alpha), e\right)* (\beta, g) & = \left(\Pi(\alpha) \Pi(\beta), \; g\right)\\
  & = \left(\Pi(\beta), g\right)* (\alpha, e) = \left(\Pi(\beta)
\Pi(^g \alpha), \; g \right),
\end{align*}
and, therefore,
\begin{equation}\label{3.1}
\pi(\alpha)\cdot \Pi(\beta) = \Pi(\beta) \cdot
\Pi(^g{\alpha}).
\end{equation}
  
Now Eve also knows Bob's public key given by
\begin{equation}\label{3.2}p_{B} = (n_b, e) * (b_1, g_{b_1}) * \cdots * (b_\ell, g_{b_\ell}) = (n_b, e) * (\beta, g)
  = \left(n_b \Pi(\beta), g\right).
\end{equation}
Combining \eqref{3.1} and
\eqref{3.2} Eve obtains
\[
\Pi(\alpha)\cdot n_b^{-1} \cdot p_B =
n_b^{-1}\cdot p_B \cdot \Pi(^g \alpha),
\]
which may be rewritten as
\begin{equation}\label{3.3}
n_b \cdot \Pi(\alpha) = p_B\cdot \Pi\left(^g \alpha\right) \cdot
p_B^{-1} \cdot n_b.
\end{equation}
  
The authors of [KTT] then assume that $N$ is a subgroup of $GL(n,
\Bbb F)$ for some positive integer $n$ and some finite field $\Bbb F$,
as is done in CBKAP.  With this assumption, and the assumption
that it is possible to generate many spurious elements $(\alpha ,e)
\in A \subset M\rtimes G$, the authors show that it may be possible
for Eve to find $n_b$ by linear algebra: Eve uses the $(\alpha,e ) $
to generate many equations of the form 
\begin{equation}\label{3.4}
n_b y_i = y'_i n_b \qquad
y_{i}, y'_i \in GL(n, \Bbb F), i = 1,2,3,\ldots.
\end{equation}
With many such equations she can then try to solve 
for $n_b.$

\section{Defeating the Kalka-Teicher-Tsaban Attack} \label{defeat}

We now describe how the TTP can choose data so that Alice and Bob can
thwart Eve's attack.  The key is to take more care in choosing the
matrix $m\in GL (n,\F)$ that is used to construct the monoids $N_{A}$,
$N_{B}$.

First, the TTP chooses $E$-commuting submonoids $A, B$ by giving a set
of generators for each of these monoids.
  
Next, the TTP chooses an element $(\beta, 1)$ out of the generators of
$B$, chooses constants $c_\ell \in \Bbb F$, and defines a matrix $$m =
\sum_\ell c_\ell \cdot \Pi(\beta)^\ell.$$ This matrix $m$ is made
public.

Then the TTP defines $N_A = N_B = \Bbb F[m]$ to be the set of all
polynomials in $m$ with coefficients in $\Bbb F.$ These two submonoids
clearly commute with each other.  Alice and Bob then choose first
private keys $n_a, n_b$ by choosing polynomials in the matrix
$m$. 

We claim that this defeats the attack.  Indeed, suppose Bob chooses
$n_b = \sum_\ell \nu_\ell m^\ell$ with $\nu_\ell \in \Bbb F$.  This
$n_{b}$ will be a solution to all the equations of the form
\eqref{3.3} and \eqref{3.4} that Eve can generate.  But this does not
give much information about $n_{b}$, since it is clear that any matrix
of the form $$n_b \cdot \sum_\ell w_\ell \cdot m^\ell, \qquad w_\ell
\in \Bbb F,$$ will also be a solution to \eqref{3.3} and \eqref{3.4}
for any choice of $w_\ell \in \Bbb F$. In general this is such a large
collection of matrices that the equations \eqref{3.3} and \eqref{3.4}
give no useful information.  Thus Eve cannot feasibly recover Bob's
first private key via this attack.
 
\begin{remark}
There is a variant protocol that deserves mention, in which the TTP
chooses commuting monoids $A,B$ and gives $B$ to Bob and only makes
$A$ public. Thus $B$ is kept secret and is only known to Bob.  The TTP
also creates the matrix $m$ out of a spurious element $(\beta, 1)$ in
$B$ as above, and makes $m$ public. Using $A$ and the matrix $m$,
Alice can do a key exchange with Bob.  This protocol is what is used
in potential RFID applications, cf.~\cite[\S 1.4]{stevens} and \cite{aggeg}.
\end{remark}
\bibliographystyle{amsplain}
\bibliography{paper}

\providecommand{\bysame}{\leavevmode\hbox to3em{\hrulefill}\thinspace}
\providecommand{\MR}{\relax\ifhmode\unskip\space\fi MR }
% \MRhref is called by the amsart/book/proc definition of \MR.
\providecommand{\MRhref}[2]{%
  \href{http://www.ams.org/mathscinet-getitem?mr=#1}{#2}
}
\providecommand{\href}[2]{#2}
\begin{thebibliography}{1}

\bibitem{ae}
I.~Anshel, M.~Anshel, D.~Goldfeld, and S.~Lemieux, \emph{Key agreement, the
  {A}lgebraic {E}raser{$^{\rm TM}$}, and lightweight cryptography}, Algebraic
  methods in cryptography, Contemp. Math., vol. 418, Amer. Math. Soc.,
  Providence, RI, 2006, pp.~1--34.

\bibitem{aggeg}
I.~Anshel, D.~Goldfeld, and P.~E. Gunnells, \emph{Fast asymmetric encryption
  using the {Algebraic Eraser}}, in preparation.

\bibitem{stevens}
P.~E. Gunnells, \emph{On the cryptanalysis of the generalized simultaneous
  conjugacy search problem and the security of the algebraic eraser}, 2011,
  arXiv:1105.1141.

\bibitem{ktt}
A.~Kalka, M.~Teicher, and B.~Tsaban, \emph{{Cryptanalysis of the Algebraic
  Eraser and short expressions of permutations as products}}, 2008,
  arXiv:0804.0629v4.

\end{thebibliography}

\end{document}